\documentclass[12pt]{iopart}

\usepackage{graphicx}
\usepackage{xcolor}
\usepackage{cite}
\usepackage{color}
\usepackage{hyperref}

\begin{document}
	
\title{Many-body localization on finite generation fractal lattices}
	
\author{Sourav Manna$^{1,2,*}$, B\l{}a\ifmmode \dot{z}\else \.{z}\fi{}ej Jaworowski$^{3,*}$ and Anne E. B. Nielsen$^{1,3}$}
\address{$^1$Max-Planck-Institut f\"ur Physik komplexer Systeme, D-01187 Dresden, Germany}
\address{$^2$Department of Condensed Matter Physics, Weizmann Institute of Science, Rehovot 7610001, Israel}
\address{$^3$Department of Physics and Astronomy, Aarhus University, DK-8000 Aarhus C, Denmark}	
\address{$^*$These authors contributed equally to this work}	
	
\begin{abstract}
We study many-body localization in a hardcore boson model in the presence of random disorder on finite generation fractal lattices with different Hausdorff dimensions and different local lattice structures. In particular, we consider the Vicsek, T-shaped, Sierpinski gasket, and modified Koch-curve fractal lattices. In the single-particle case, these systems display Anderson localization for arbitrary disorder strength if they are large enough. In the many-body case, the systems available to exact diagonalization exhibit a transition between a delocalized and localized regime, visible in the spectral and entanglement properties of these systems. The position of this transition depends on the Hausdorff dimension of the given fractal, as well as on its local structure.
\end{abstract}

\section{Introduction}

The interplay between disorder and interactions in closed quantum many-body systems gives rise to transitions between ergodicity and many-body localization \cite{nandkishore1}. Many-body localized systems are nonergodic and do not obey the eigenstate thermalization hypothesis. This phenomenon goes beyond the ground state properties and involves the physics of highly excited states. The transitions have been studied theoretically \cite{huse2, PhysRev.109.1492,PhysRevLett.95.206603, PhysRevB.75.155111, PhysRevB.77.064426, PhysRevLett.117.027201} and experimentally \cite{schreiber1,PhysRevLett.114.083002,choi1,smith1}. The transition between ergodicity and many-body localization has also exposed rich structures in the entanglement entropy \cite{PhysRevLett.109.017202, PhysRevB.93.174202, PhysRevX.7.021013, PhysRevB.96.020408}.

The study of many-body localization has been extensively carried out in one-dimensional systems \cite{huse2,PhysRev.109.1492,PhysRevLett.95.206603,PhysRevB.77.064426, PhysRevB.75.155111,PhysRevLett.117.027201}, and the prospect of many-body localization has also been studied in two-dimensional systems \cite{choi1,wahl1, PhysRevLett.126.180602}. The existence of a many-body localized phase in the thermodynamic limit is debated \cite{PhysRevB.93.014203, PhysRevLett.117.027201, PhysRevE.102.062144, abanin2021distinguishing}. While the thermodynamic limit is important for condensed matter systems, the progress of quantum simulator technologies has made it possible to build small and highly controllable systems that can behave differently from the same system in the thermodynamic limit \cite{blume2012few}. For these few-particle systems, relevant for experiments with cold atoms, the existence of a many-body localized regime is well established experimentally and theoretically both in one dimension \cite{huse2,PhysRevB.75.155111,schreiber1} and two dimensions \cite{choi1,wiater2018impact}.

Fractal lattices are aperiodic structures which exhibit fractional dimensions. Such lattices are playgrounds to explore properties, which depend on the dimension of space and how the particles can move. They provide an intermediate regime between integer dimensions (e.g.\ between one and two, or two and three). The fractals also display peculiar properties which are absent in integer dimensions, such as the lack of distinction between the bulk and edge. Several physical properties were studied in these systems \cite{PhysRevE.91.012118,van2016quantum,westerhout2018plasmon, brzezinska2018topology,PhysRevE.98.062114,iliasov2019power, pai2019topological,fremling2019chern,Manna3,Manna2}. In particular, single-particle (Anderson) localization phenomena were investigated in fractal and bifractal setups \cite{rammal1983random,schreiber1996dimensionality, travenec2002distribution,kosior2017localization}. The possibility of many-body localization in systems other than periodic lattices, was in general rarely studied (an example is \cite{balasubramanian2020many}, where the formalism was developed for models defined on any graph). The studies of fractal lattices are also motivated by experimental developments to generate fractals \cite{shang2015assembling,kempkes2019design}, including the possibility to realize arbitrary lattices with Rydberg atoms \cite{barredo2016atom} and the progress in techniques of constructing arbitrary lattice geometries and addressing single sites in optical lattices \cite{bakr2009quantum,kuhr2010single}.

Here, we investigate models of hardcore bosons on finite generation fractal lattices in the presence of disorder and compare the results to one- and two-dimensional systems. The considered fractal lattices have different local structures and different Hausdorff dimensions, and we generally find that the systems transition from an ergodic to a many-body localized regime as the disorder strength is increased. Our results suggest that the strength at which the transition happens is influenced by several factors, but typically increases with increasing Hausdorff dimension. The transitions are not sharp phase transitions, as the considered lattices are far from the thermodynamic limit, and we stress that our results do not say whether the many-body localized phase exists in the thermodynamic limit or not. Extrapolating to the thermodynamic limit is challenging due to the rapid increase of the Hilbert space dimension and is not realistic with the exact diagonalization tools used here. Our results instead show how small, experimentally relevant systems are affected by disorder. We also briefly discuss the single-particle case, which shows Anderson localization.

The paper is organized as follows. We describe the fractal lattices and the hardcore boson model in section \ref{lattices}. In section \ref{sec:anderson}, we study the single-particle localization in these systems. In section \ref{sec:jordanwigner}, we discuss the relation between the many-particle hardcore boson model and fermionic models. The spectral and entanglement properties of the many-body systems, showing a transition from ergodic to many-body localized behavior, are investigated in sections \ref{spectral} and \ref{EE}, respectively. We conclude the paper in section \ref{concl}.

\section{Fractal lattices and the model}\label{lattices}
\begin{figure*}
\begin{indented}\item[]
\includegraphics[width=\linewidth]{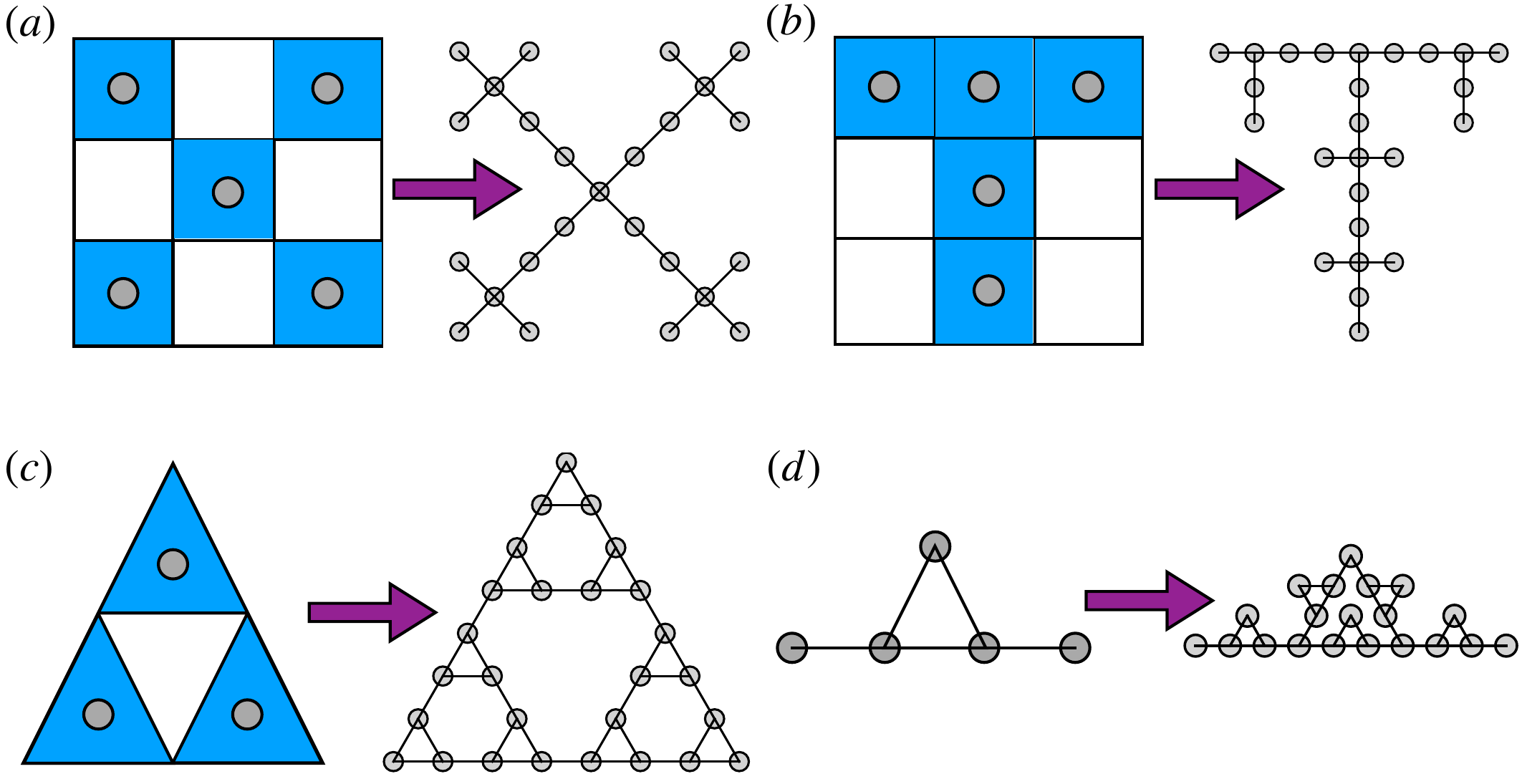}
\caption{We consider different fractal lattices, namely (a) the Vicsek fractal lattice, (b) the T-shaped fractal lattice, (c) the Sierpinski gasket fractal lattice, and (d) a modified Koch-curve fractal lattice. To the left we show the first generation of the fractals, and the lattice sites are illustrated as grey circles. To the right we show either the second or the third generation of the lattices, and the allowed nearest-neighbor hopping terms are shown with black, solid lines. Note that in (d) we only keep those nearest-neighbor hoppings, which maintain the self-similarity of the fractal lattice. The number of lattice sites is (a-b) $N=25$, (c) $N=27$, and (d) $N=20$, respectively.}\label{lat}
\end{indented}
\end{figure*}

In the following, we consider fractal lattices with different Hausdorff dimensions and different local lattice structures, and we start with a description of these lattices. The fractal lattices are shown in figure \ref{lat}.

The Vicsek fractal lattice is constructed from a square as follows. We first divide the square into $3 \times 3$ squares. Then the four squares at the corners and the square in the middle are kept and the other squares are removed. Hence the remaining squares form a $``\times"$ shape. We repeat the procedure recursively for each of the five remaining squares to obtain different generations. Once the desired generation has been reached, a lattice site is placed on each of the remaining squares to obtain the fractal lattice. The Vicsek fractal lattice has a fractal dimension $\mathcal{D}=\ln(5)/\ln(3) \approx 1.465$. A second generation lattice with $N=25$ sites is shown in figure \ref{lat}(a).

The construction of the T-shaped fractal lattice is the same as for the Vicsek fractal lattice, except that we keep the squares forming a ``T" shape. The T-shaped fractal lattice has fractal dimension $\mathcal{D}=\ln(5)/\ln(3) \approx 1.465$. A second generation lattice with $N=25$ sites is shown in figure \ref{lat}(b).

The Sierpinski gasket fractal lattice is constructed from an equilateral triangle by repeated removal of triangular subsets as follows. We begin with an equilateral triangle, which we subdivide into four smaller congruent equilateral triangles. Then we remove the central triangle. We repeat this step with each of the remaining smaller triangles to obtain different generations, and finally we obtain the fractal lattice by putting one lattice site on the center of each of the remaining triangles. The Sierpinski gasket fractal lattice has fractal dimension $\mathcal{D}=\ln(3)/\ln(2) \approx 1.585$. A third generation lattice with $N=27$ sites is shown in figure \ref{lat}(c).

The modified Koch-curve fractal lattice is constructed as follows. We begin with a straight line and divide the line segment into three segments of equal length. We draw an equilateral triangle that has the middle segment from the previous step as its base. We place one lattice site at each of the vertices and we repeat the procedure to have different generations. The modified Koch-curve fractal lattice has a fractal dimension $\mathcal{D}=\ln(5)/\ln(3) \approx 1.465$. A second generation lattice with $N=25$ sites is shown in figure \ref{lat}(d).

The fractals in figure \ref{lat}(a), (b), and (d) all have the same fractal dimension but have different local lattice structures. For comparison, we also consider a one-dimensional chain and a two-dimensional square lattice with open boundary conditions.

On the fractal lattices, as well as on the one- and two-dimensional lattices, we define a hardcore boson model. Its Hamiltonian is
\begin{equation}\label{dis_ham}
H= -\sum_{\langle ij \rangle} c_i^\dagger c_j + \sum_{i} h_i n_i,
\end{equation}
where $c_i$ is the annihilation operator of a hardcore boson, acting on the $i$th lattice site, and $n_i=c_i^\dag c_i$. The first sum is over the nearest neighbor sites of the lattice that are connected with bonds in figure \ref{lat} (or, in the case of one- and two-dimensional lattices, over nearest-neighbor sites). Note that each bond contributes two terms making the Hamiltonian Hermitian. We take $h_i \in [-h,h]$ to be random variables, which are uniformly distributed in the interval from $-h$ to $h$, where $h$ is the disorder strength. This model is motivated by recent experiments \cite{greiner1} in quasi-one-dimensional optical lattices.

\section{Single-particle properties}\label{sec:anderson}

\begin{figure}
\begin{indented}\item[]
\includegraphics[width=\linewidth]{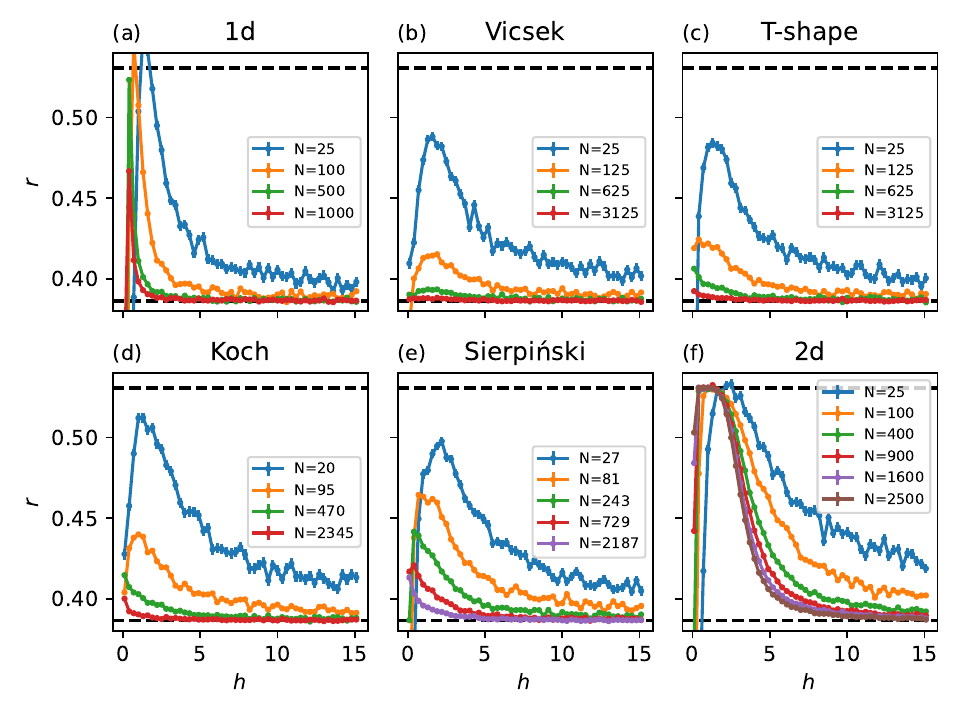}
\caption{The average level spacing ratio $r$ for the single-particle systems as a function of disorder strength $h$. The dashed horizontal lines are located at $r\approx 0.386$ and $r\approx 0.536$, corresponding to Poisson and Wigner-Dyson statistics, respectively.}
\label{fig:r-anderson}
\end{indented}
\end{figure}

Before delving into the physics of many-particle systems, we study the single-particle localization problem on the fractal lattices. That is, we consider the Hamiltonian \eref{dis_ham}, but fix the number of particles to be $M=1$. In this case, the interaction enforced by the hardcore condition, as well as the statistics of the particles, have no effect, and the problem is equivalent to the single-particle free fermion problem.

In infinite one- and two-dimensional systems, arbitrarily weak disorder is enough to localize the wavefunctions \cite{abrahams1979scaling}. In finite systems, the localization is seen when the system size is large compared to the localization length. Here, we consider the fractals from section \ref{lattices} to see if their behavior is similar. To probe the localization of the single-particle wavefunctions, we study the level spacing statistics and inverse participation ratio. We average over 1000 disorder realizations.

\textit{Level spacing statistics.--}
In the ergodic phase of systems with time-reversal symmetry, the statistical distribution of the energy level spacings obeys the Wigner-Dyson surmise of the Gaussian orthogonal ensemble as the energy levels exhibit level repulsion. On the contrary the Poisson distribution is expected in the localized phase. Therefore to discriminate between these two behaviors we consider the ratio of consecutive level spacings \cite{tikhonov2016anderson,PhysRevB.75.155111,huse2,bhatt1,laumann1} defined as
\begin{equation}
r_n = \frac{\min(\delta_n,\delta_{n+1})}{\max(\delta_n,\delta_{n+1})},\qquad \delta_n = E_{n+1}-E_n,
\end{equation}
where the energies $E_n$ of the eigenstates are labeled in increasing order with the index $n\in\{1,2,\ldots,{dim} \mathcal{H}\}$ and ${dim} \mathcal{H}$ is the dimension of the Hilbert space (for $M=1$, ${dim} \mathcal{H}=N$). The value $r = \langle \bar{r}_n \rangle$ represents the average of $\bar{r}_n$ over disorder realizations, and the value $\bar{r}_n$ represents the average
of $r_n$ over the middle $1/3$ portion of the energy spectrum. The value of $r$ approaches $r=4-2\sqrt{3} \approx 0.536$ in ergodic systems with time-reversal symmetry and $r = 2 \ln(2) -1 \approx 0.386$ in localized systems \cite{roux1}.

\begin{figure}
\begin{indented}\item[]
\includegraphics[width=\linewidth]{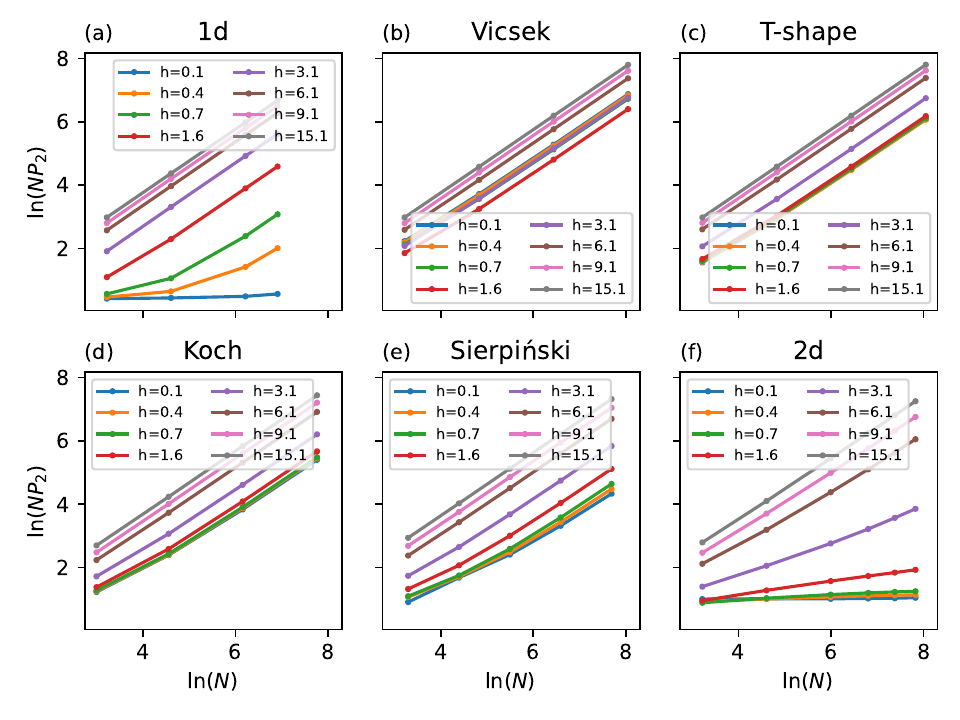}
\caption{The average inverse participation ratio for the single-particle systems, for different disorder strengths, as a function of system size.}
\label{fig:ipr-anderson}
\end{indented}
\end{figure}

The results are shown in figure \ref{fig:r-anderson}. All the fractals seem to display a behavior qualitatively similar to the behavior of the one-dimensional system. For small systems and weak disorder, $r$ has a value significantly higher than $0.386$, but not necessarily close to $0.536$, suggesting a delocalized regime. With increasing disorder strength, $r$ gets close to $0.386$, which suggests localization. As we increase the system size, the maximum of $r$ decreases, and the entire curve gets closer and closer to $0.386$, suggesting that in the thermodynamic limit, the system will be localized for any disorder strength. For the two-dimensional system, we observe a delocalized regime with $r \approx 0.536$ at weak disorder. The position of the transition shifts toward lower $h$, however, as the system size increases, which is consistent with the theoretical result that in the infinite two-dimensional system localization occurs for an arbitrarily weak disorder.

\textit{Inverse participation ratio.--} For the $n$th eigenstate of the single-particle Hamiltonian, the inverse participation ratio is defined as
\begin{equation}
P_2(n)=\sum_{i}|\psi_i|^4,
\end{equation}
where $\psi_i$ is the coefficient of the eigenstate at site $i$. The inverse participation ratio measures the degree of localization. If a wavefunction is spread evenly over the sites, we have $P_2(n)=1/N$, while if it is fully localized on a single site, $P_2(n)=1$ (see \cite{tikhonov2016anderson} for an example of how the inverse participation ratio is used in an Anderson localization problem). Thus, in the ergodic system, $P_2(n)$ is inversely proportional to the number of sites, while in the localized system it is constant. Similarly to $r$, we define $P_2=\langle \overline{P_2(n)} \rangle$ as the value of $P_2(n)$ averaged over the middle 1/3 of the spectrum and over disorder realizations.

In figure \ref{fig:ipr-anderson}, we plot $\ln(NP_2)$ as a function of $\ln(N)$. For strong disorder, the dependence is seen to be almost linear with slope close to 1, which signifies localization. For weak disorder, the dependence becomes less linear, but otherwise roughly similar, except of two cases: the one- and two-dimensional systems at weak disorder. In these cases, for $h=0.1$ (the weakest investigated disorder) we observe that $\ln(NP_2)$ is almost constant.

The results from figures \ref{fig:r-anderson} and \ref{fig:ipr-anderson} suggest that the studied fractals indeed behave as an interpolation between one and two dimensions, in the sense that they seem to display localization for arbitrarily weak disorder, provided the system size is large enough. Moreover, it seems that the fractals do not display the $P_2\sim 1/N$ regime even at $h=0.1$. This suggests that the obstructions to localization in small systems may be less pronounced in the fractal lattices. An alternative explanation is that the fractal structures are not uniform, so there is no reason for the wavefunction to spread evenly over the sites. We note that very pronounced inhomogenities were observed on the Cayley trees (finite Bethe lattice), which have a large fraction of edge sites. In such a case, it was possible for localized states to occur near the edges while delocalized states exist near the center of the lattice in finite systems \cite{tikhonov2016fractality,sonner2017multifractality}. We do not rule out similar effects in our tree-like fractals (Vicsek, T-shaped), but we note that our systems are more complicated: there are more than two types of sites with different local environments. Thus, we focus on the quantities averaged over $1/3$ of the spectrum. We also note that there are significant differences between these fractals and Cayley trees.

\begin{figure}
\begin{indented}\item[]
\includegraphics[width=\linewidth]{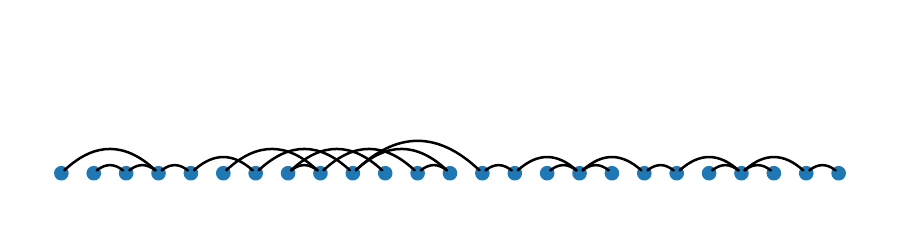}
\caption{One of the possible ways of folding the T-shaped fractal (figure \ref{lat}(b)) into a one-dimensional chain, with black lines denoting the hoppings. After the Jordan-Wigner transformation, the nearest-neighbor hoppings become single-particle fermionic hoppings, while the more distant hoppings transform into hoppings with phases depending on the occupation of all intermediate sites.}\label{fig:jordanwigner}
\end{indented}
\end{figure}

\section{Jordan-Wigner mapping for many-particle systems}\label{sec:jordanwigner}

Since the $M=1$ case is equivalent to the problem of free fermions, it is instructive to map our hardcore bosons to fermions using the Jordan-Wigner transformation. In order to do it, we have to number the lattice sites from $1$ to $N$, which can also be seen as a mapping to a one-dimensional chain with hoppings that are generally non-local. The Jordan-Wigner transformation transforms the hopping $-c_i^{\dagger}c_j+h.c.$ between sites $i$ and $j$, with $i<j$, into
\begin{equation}
-c_i^{\dagger}(-1)^{\sum_{k=i}^{j-1}n_k}c_j+h.c.
\end{equation}
Thus, the nearest-neighbor hoppings of bosons turn into nearest-neighbor hoppings of fermions, but $l$th-neighbor hoppings ($l=i-j$) turn into interaction terms involving up to $l$ particles.

The simplest case is the one-dimensional system, which transforms to a chain of free fermions. The two-dimensional case is more complicated, but it has a natural way of ordering the sites when mapping to a one-dimensional chain, which leads to interactions involving up to five particles for the $5\times 5$ system considered in this work. For fractal lattices, there is no obvious way of ordering the sites (and e.g.\ for 25 sites there are more than $10^{25}$ permutations). One example for a T-shaped fractal is shown in figure \ref{fig:jordanwigner}. It can be seen that the system maps into several free fermion chains of different lengths (the sites connected by nearest-neighbor hoppings in figure \ref{fig:jordanwigner}), connected by interaction terms involving several particles (more distant hoppings in figure \ref{fig:jordanwigner}). We expect a similarly complicated result for other fractals as well.

Therefore, the Jordan-Wigner transformation simplifies the problem only in the one-dimensional case. The one-dimensional system, being equivalent to free fermions, is integrable for any value of the disorder and is Anderson localized for the disorder strong enough that the localization length is smaller than the system size (see figure \ref{fig:ipr-anderson}(a)). For the two-dimensional case we can rely on previous experiments and calculations, showing that finite systems of hardcore bosons exhibit a transition between ergodic and many-body localized regimes \cite{wahl1, PhysRevLett.126.180602, choi1}. The fractal lattices will be studied numerically in the following sections.

\begin{table}
\begin{indented}\item[]
\begin{tabular}
{ p{3.4cm} p{1cm} p{0.6cm} p{0.6cm} p{1cm} p{1.0cm}}
\hline\hline
Lattices & $\mathcal{D}$ & $N$ & $M$ & $M/N$ &${dim} \mathcal{H}$ \\ \hline
One-dimensional & 1 & $25$ & $4$ &$0.16$ & $12650$ \\
Vicsek & 1.465 & $25$ & $4$ &$0.16$ & $12650$ \\
T-shaped & 1.465 & $25$ & $4$ &$0.16$ & $12650$ \\
Koch-curve like & 1.465 & $20$ & $3$ &$0.15$ & $\phantom{1}1140$\\
Sierpinski gasket & 1.585 & $27$ & $4$ &$0.148$ & $17550$ \\
Two-dimensional & 2 & $25$ & $4$ &$0.16$ & $12650$ \\
\hline\hline
\end{tabular}
\caption{The Hausdorff dimension $\mathcal{D}$, the number of lattice sites $N$, the number of particles $M$, the lattice filling factor $M/N$, and the corresponding Hilbert space dimension ${dim} \mathcal{H}$ for the considered many-body systems.}
\label{tab-1}
\end{indented}
\end{table}

\begin{figure}
\begin{indented}\item[]
\includegraphics[width=\linewidth]{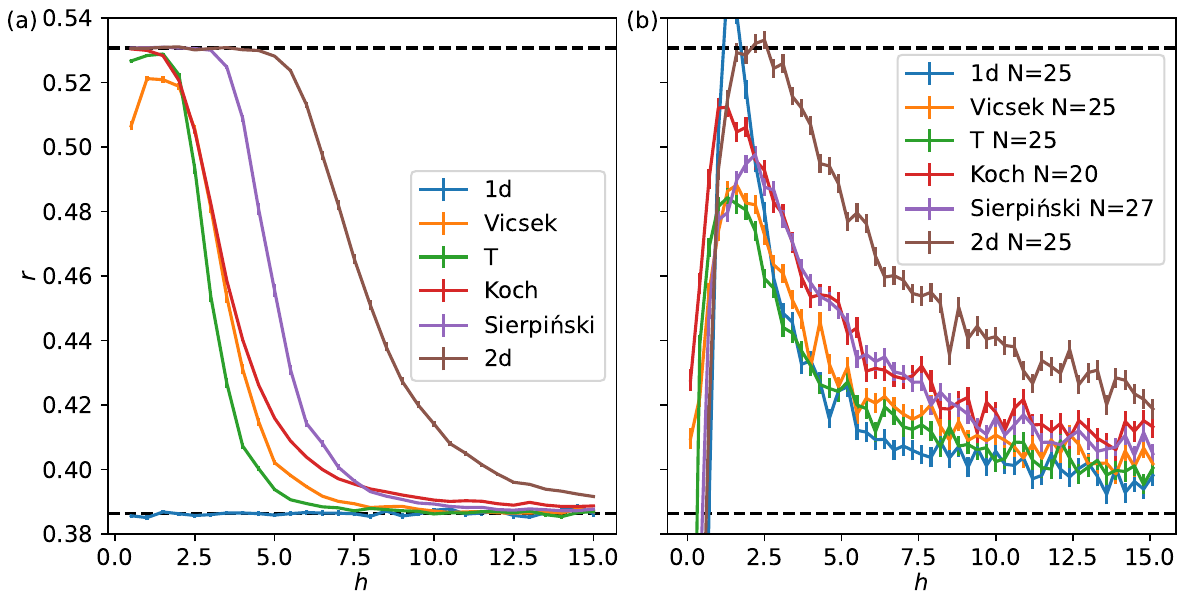}
\caption{(a) The adjacent gap ratio $r$ as a function of the disorder strength $h$ for the considered lattices in the many-particle case. The dotted horizontal black lines at $r \approx 0.536$ and at $r \approx 0.386$ are the expected values for ergodic and many-body localized systems, respectively. For all the cases, there is a transition from ergodic behavior to many-body localized behavior as the disorder strength is increased. (b) The adjacent gap ratio $r$ as a function of the disorder strength $h$ for the same systems as in (a), but with one particle only ($M=1$).} \label{SS}
\end{indented}
\end{figure}

\section{Many-body spectral properties}\label{spectral}

We now use exact diagonalization to determine the properties of model (\ref{dis_ham}) on the finite fractal lattices for $M>1$. Due to the rapid increase of the Hilbert space dimension with system size, we limit the study to the system sizes listed in table \ref{tab-1}. We choose the particle number in such a way that the filling factor $M/N$ is nearly the same in all the lattices. We differentiate between thermal and many-body localized behaviors by studying the level spacing statistics and the many-body mobility edge with tools from random matrix theory.

\textit{Level spacing statistics.--}
The level spacing statistics are investigated using the parameter $r$, defined in section \ref{sec:anderson}. In figure \ref{SS}(a), we plot $r$ as a function of the disorder strength $h$. The value of $r$ is obtained by averaging over the middle 1/3 of the states in 250 disorder realizations (except from the Koch lattice, for which we study 2500 realizations, because for this case we consider $M=3$ for which the Hilbert space is much smaller).

For the one-dimensional lattice, we observe that $r$ is close to $0.386$ for all $h$. This is related to the fact that the one-dimensional case is equivalent to a free fermion system and thus integrable. For all the other lattices, we find a transition from ergodic behavior to a many-body localized behavior. This confirms that the fractals (even the tree-like ones) cannot be mapped to a free fermion model.

The three lattices with Hausdorff dimension $1.465$ all have the transition at comparable disorder strengths. The transition happens at a slightly larger disorder strength for the lattice with Hausdorff dimension $1.585$, and for the two-dimensional lattice the transition happens at even stronger disorder. The fractal lattices hence show features intermediate between the one-dimensional and the two-dimensional case, and the Hausdorff dimension seems to be an important factor regarding the location of the transition.

A comparison between the single- and many-body results can be performed by replotting all the blue curves from figure \ref{fig:r-anderson} in one picture, which we do in figure \ref{SS}(b). We note that the many-body results display a clearer transition between a localized and delocalized regime, as $r$ for fractal lattices in figure \ref{SS}(b) never reaches the Wigner-Dyson value. Also, the Hausdorff dimension seems somewhat less important in the single-particle than in the many-particle case -- note the similarity between the $r$ versus $h$ curves for the Koch fractal ($\mathcal{D}\approx 1.465$) and Sierpinski triangle ($\mathcal{D} \approx 1.585$), as well as between the one-dimensional case and the Vicsek and T-shaped fractals ($\mathcal{D}\approx 1.465$), in figure \ref{SS}(b). The influence of the Hausdorff dimension may be obscured by the variations in the system size ($N=20$ for Koch fractal and $N=27$ for Sierpinski lattice -- the latter has 1.35 times more sites).

\begin{figure*}
\begin{indented}\item[]
\includegraphics[width=\linewidth]{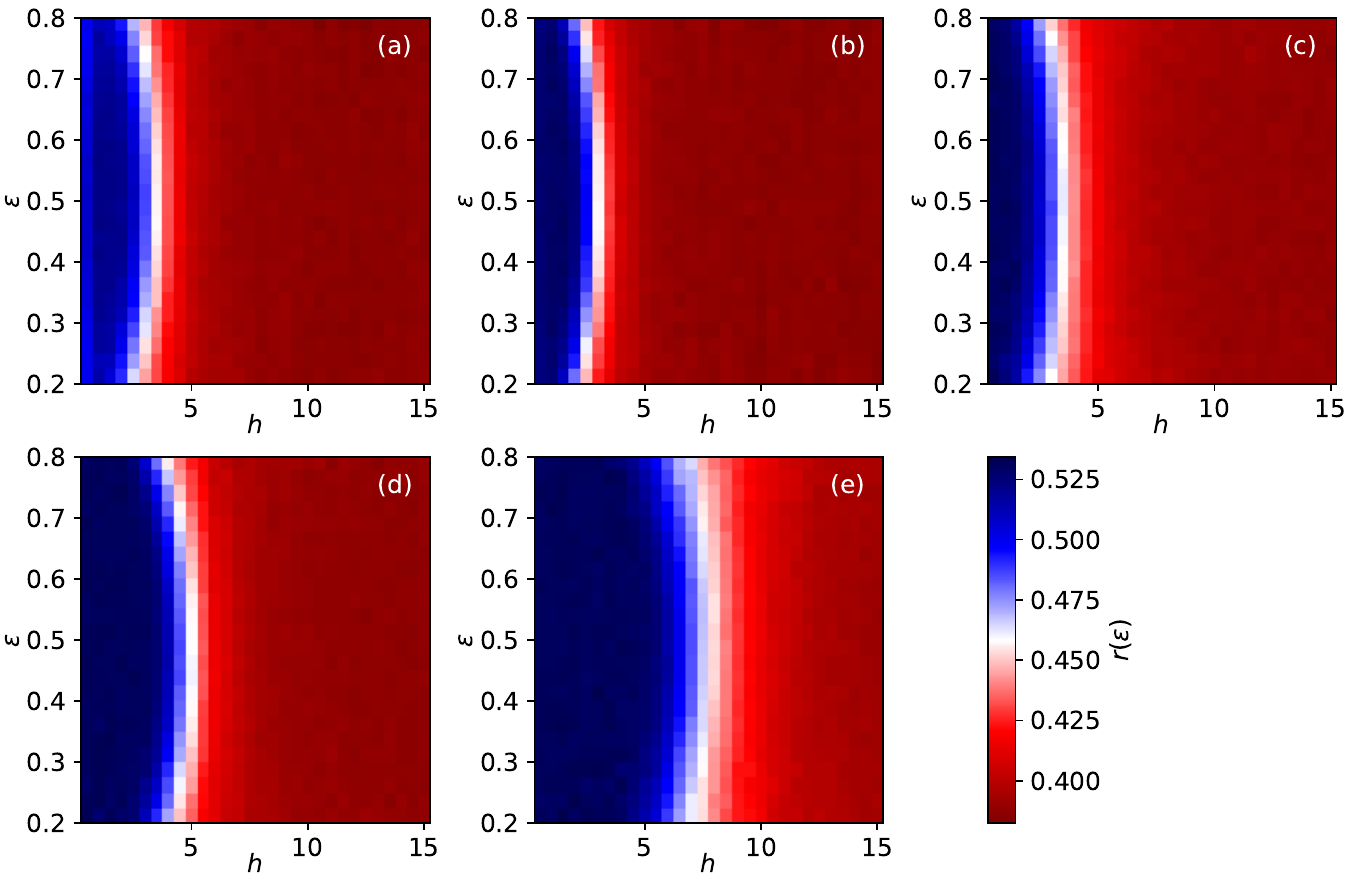}
\caption{The energy density resolved adjacent gap ratios $r(\epsilon)$ as a function of the disorder strength $h$ for (a) the Vicsek fractal lattice, (b) the T-shaped fractal lattice, (c) the modified Koch-curve fractal lattice, (d) the Sierpinski gasket fractal lattice, and (e) the two-dimensional square lattice.}\label{ME}
\end{indented}
\end{figure*}

\textit{Many-body mobility edge.--}
The many-body mobility edge is the energy densities at which there are transitions between thermal and many-body localized states \cite{Mauro1,laumann1,laumann2,sheng1,mondaini1}. For a given energy $E$, the corresponding energy density is defined as
\begin{equation}
\epsilon = \frac{E-E_{\min}}{E_{\max}-E_{\min}},
\end{equation}
where $E_{\min}$ ($E_{\max}$) is the lowest (highest) energy eigenvalue in the energy spectrum.

We plot the energy density resolved gap spacings $r(\epsilon)$  (i.e.\ $r_n$ averaged over disorder realizations as well as over the eigenstates falling into a certain small window of $\epsilon$) in the $\epsilon$ versus $h$ plane in figure \ref{ME}. For both the fractal lattices and the two-dimensional lattice a D-shape appears in the plots. In the region with the mobility edge, there are localized states at low and high $\epsilon$ and thermal states for intermediate $\epsilon$. Also in this plot it is again seen that the transition from the thermal to the many-body localized behavior happens at lower disorder strength for the models on the fractal lattices than for the model on the two-dimensional lattice.

\begin{figure*}
\begin{indented}\item[]
\includegraphics[width=\linewidth]{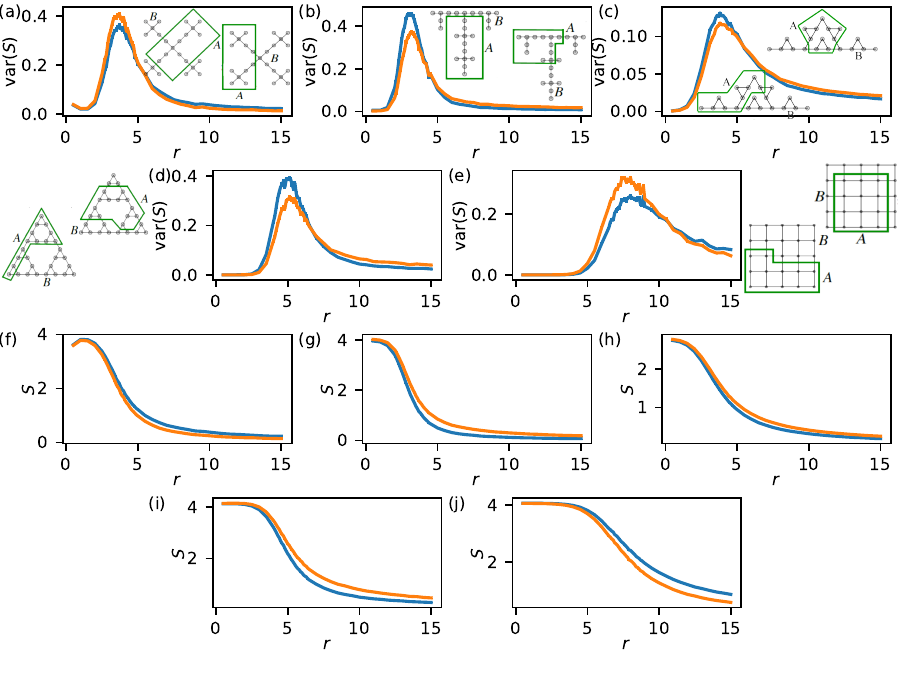}\\
\caption{The variance $\mathrm{var}(S)$ and mean $\langle\bar{S}\rangle$ of the entanglement entropy as a function of the disorder strength $h$ for (a,f) the Vicsek fractal lattice, (b,g) the T-shaped fractal lattice, (c,h) the modified Koch-curve fractal lattice, (d,i) the Sierpinski gasket fractal lattice, and (e,j) the two-dimensional square lattice. We take two choices of the subsystems $A$ and $B$ for each case as shown in the insets, and we find similar behaviors for both. The peaks in $\mathrm{var}(S)$ show the transition points from thermal behavior at weak disorder to the many-body localized behavior at strong disorder. The mean entropy $\langle\bar{S}\rangle$ is high in the thermal region and low in the many-body localized region.}\label{EE_plot}
\end{indented}
\end{figure*}

\begin{figure}
\begin{indented}\item[]
\begin{center}
\includegraphics[width=0.6\linewidth]{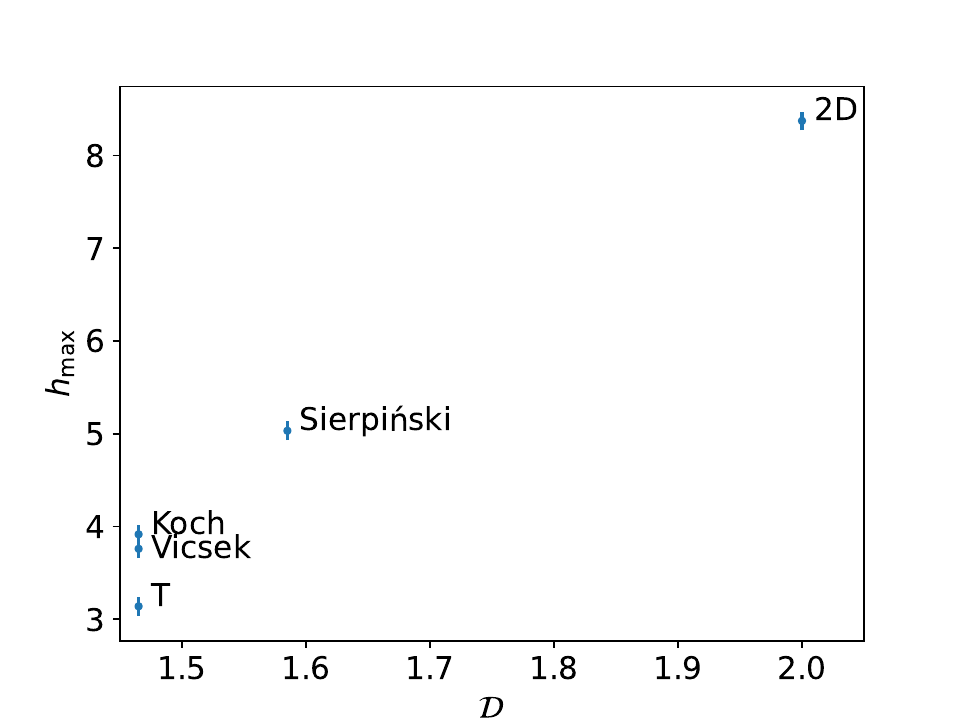}\\
\caption{The position of the maximum of the variance of the entanglement entropy in the studied lattices plotted as a function of the Hausdorff dimension.}\label{EE_varmax}
\end{center}
\end{indented}
\end{figure}

\section{Many-body entanglement entropy}\label{EE}
The entanglement entropy is another quantity to identify the transition between ergodicity and many-body localization \cite{pollman1,sheng1}. The variance of the entanglement entropy has a peak at the transition point. We study this quantity for the considered fractal lattices.

We partition the system into two parts, which we denote $A$ and $B$. The von Neumann entanglement entropy of part $A$ of the state $|\Psi \rangle$ is
\begin{equation}
S = - \mathrm{Tr}_A\left[\rho_A\ln(\rho_A)\right],
\label{eq:entropy}
\end{equation}
where
\begin{equation}
\rho_A = \mathrm{Tr}_B (|\Psi\rangle \langle \Psi|)
\end{equation}
is the reduced density matrix of part $A$ and the traces are over part $A$ and $B$, respectively. We consider $20$ energy eigenstates closest to the middle of the spectrum $\epsilon=0.5$. The mean entropy $\langle \bar{S}\rangle$ is obtained by averaging \eref{eq:entropy} over all these states and all the disorder realizations. Because we focus only on a small number of states, we can obtain them using the shift-invert approach, which is faster than full matrix diagonalization and allows us to study a larger number of disorder realizations: 10000 for the Koch lattice and 2500 for the other lattices.

The variance $\mathrm{var}(\bar{S})$ of the entanglement entropy is computed in the following way: first, we obtain the average $\bar{S}$ of the entropy over 20 states in the middle of the energy spectrum in each disorder realization, and then compute the variance $\mathrm{var}(\bar{S})$ of that average among the disorder realization. We plot the variance $\mathrm{var}(\bar{S})$ and the mean $\langle \bar{S}\rangle$ of the entanglement entropy as a function of the disorder strength $h$ in figure \ref{EE_plot} for all the lattices. We note that the choices of $A$ and $B$ are not unique on fractal lattices and hence we make two different choices in each case for comparison. The values of $\mathrm{var}(\bar{S})$ are small both deep into the ergodic region and deep into the many-body localized region but are large in the vicinity of the transition point. The mean value of the entropy $\langle \bar{S}\rangle$ is high in the thermal region and low in the many-body localized region.

For a quantitative comparison of the transition in different lattices, we study the position of the maximum of $\mathrm{var}(\bar{S})$. To estimate the error bar on the maximum, we apply the following procedure. We divide the disorder realization into blocks of size 100. For each block we compute the variance $\mathrm{var}(\bar{S})_{\mathrm{block}}$ as a function of $h$. Then, we find the value $h_{\max}$ corresponding to the maximum of $\mathrm{var}(\bar{S})_{\mathrm{block}}$. Finally, we average $h_{\max}$ over all the blocks, obtaining the mean $\langle h_{\max} \rangle$ and the standard deviation $\sigma_{\max}$ of the mean. We use $\max\{\sigma_{\max}, 0.1\}$ as an error bar, because 0.1 is the separation between the data points in the first axes of the plots in figure \ref{EE_plot} (note that this separation is 0.5 in most parts of the plots, but we use a finer grid near the maxima of $\mathrm{var}(\bar{S})$).

The resulting transition points $\langle h_{\max} \rangle$ are shown in figure \ref{EE_varmax} as a function of the Hausdorff dimension. Both choices of bipartition yield the same results, and thus we present the plot only for one of them. A probable reason is the correlation between the results, caused by the fact that we use the same disorder realizations for both bipartitions (i.e.\ for a given disorder realization, we diagonalize the Hamiltonian and use its eigenstates to compute entropies for both bipartitions).

The transition points $\langle h_{\max} \rangle$ from figure \ref{EE_varmax} fulfil the following condition: if we take two lattices with fractal dimensions $\mathcal{D}_1$ and $\mathcal{D}_2$ and transition points $\langle h_{\max} \rangle_1$ and $\langle h_{\max} \rangle_2$, respectively, then if $\mathcal{D}_1<\mathcal{D}_2$, then $\langle h_{\max} \rangle_1 < \langle h_{\max} \rangle_2$, suggesting the importance of Hausdorff dimension in the description of many-body localization in fractals. On the other hand, if $\mathcal{D}_1=\mathcal{D}_2$, it is not necessarily the case that $\langle h_{\max} \rangle_1 = \langle h_{\max} \rangle_2$, suggesting the importance of other factors, such as local lattice structure.

\section{Conclusions}\label{concl}

We have studied the transition between ergodicity and many-body localization in a model of hardcore bosons in the presence of disorder on finite fractal lattices with Hausdorff dimensions between one and two and different local lattice structures. In particular, we have studied models on the Vicsek fractal lattice, the T-shaped fractal lattice, the Sierpinski gasket fractal lattice, and the modified Koch-curve fractal lattice. We considered a one-dimensional chain and a two-dimensional square lattice for comparison. We also compared the results to the single-particle case, where much larger system sizes are computationally available.

For the single-particle case, the results suggest Anderson localization for all disorder values, as is known for one and two dimensions. In the many-body case, a transition from an ergodic regime to a many-body localized regime was found on all the fractal lattices by studying the level spacing statistics, the many-body mobility edge, and the von Neumann entropy of highly excited states. The analysis of the variance of the entanglement entropy suggests that the disorder strength at the transition point generally increases with the Hausdorff dimension of the lattice, but is also affected by the local lattice structure.

The fractal lattices studied here provide further examples of systems beyond one dimension, which exhibit a many-body localized regime for finite size. At the same time, the number of sites scales more slowly with linear system size than in two dimensions. The fractal lattices considered here are bigger in the sense of linear size than two-dimensional lattices with the same number of sites.

In recent years, there has been developments in the field of many-body localization in two dimensions. An avalanche theory for many-body localization \cite{PhysRevLett.121.140601, PhysRevB.99.134305, PhysRevLett.125.155701} has been proposed, within which the many-body localization in two dimensions has been analyzed both analytically and numerically. It is concluded that a transition exists between ergodicity and many-body localization in a finite two-dimensional system. The disorder strength, at which the transition happens, increases, however, indefinitely with increasing system size. Hence we expect the transition disorder strength of our results in two-dimensions to be a growing function of the system size. Whether similar conclusions hold for fractal lattices remains an interesting open question.

\section{Acknowledgments}
This work has been supported by Danmarks Frie Forskningsfond under Grant No.\ 8049-00074B and by Carlsbergfondet under Grant No.\ CF20-0658. S.M.\ thanks Weizmann Institute of Science, Israel Deans fellowship through Feinberg Graduate School for financial support.

\section*{References}
\bibliographystyle{iopart-num}
\bibliography{bibfile}
	
\end{document}